\documentclass[twocolumn,showpacs,prl]{revtex4}
\usepackage{graphicx}
\usepackage{dcolumn}
\usepackage{amsmath}
\usepackage{latexsym}

\begin{document}

{\bf Comment on "Memory Effects in an Interacting Magnetic Nanoparticle System"
 by Sun {\em et. al}, \prl, 91, 167206 (2003) }\\
In a recent letter Sun {\em et. al.}\cite{sun} have presented interesting 
experimental results of history dependent magnetization 
in a monolayer of sputtered permalloy (Ni$_{81}$Fe$_{19}$) 
clusters on a SiO$_2$ substrate. 
Starting at a high (room) temperature (T = T$_{\infty}$),
the system is steadily cooled in a small magnetic field H and the magnetization 
M measured as a function of T. Subsequently at low temperatures the cooling is 
arrested at a few T-steps where H is first switched 
off and then restored after a wait of a few hours. From the lowest T
the system is heated and M(T) measured in the presence of H. 
The heating path surprisingly shows wiggles in M(T) 
at all the T-steps where H was switched off during cooling, apparently keeping 
a memory of the temperature arrests!
The authors attribute this to aging and concomitant memory-dependent effects 
found in the spin glass phase.
\vskip .1cm

\begin{figure}[h]
\begin{center}
\includegraphics[width=5.7cm,angle=270]{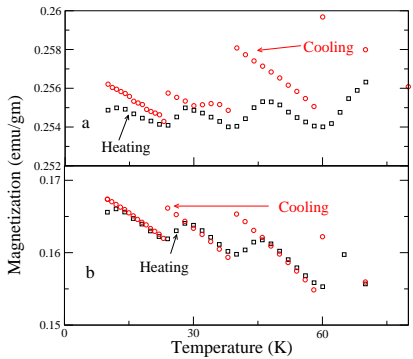}
\end{center}
\caption{Experimental M(T) curves during cooling (red) and heating 
(black) for the (a) interacting and (b) non-interacting cases}
\label{expt}
\end{figure}

\noindent
In this Comment we report (Fig.\ref{expt}) an identical phenomenon in 
systems of NiFe{$_2$}O{$_4$} particles embedded in a 
SiO{$_2$} matrix with two different interparticle spacings 4 nm (1) and
15 nm (2),  
which controls the strength of the dipolar interactions. Not only 
do we find the memory effect to be present in the non-interacting sample (2)
(Fig.\ref{expt}(b)), indeed we find it to be {\em more} prominent than in the 
interacting case (1) (Fig.\ref{expt}(a)).
\vskip .1cm

\noindent 
We demonstrate that the phenomenon observed in Ref.\cite{sun} and shown in 
Fig.\ref{expt} can be simply attributed to a 
superposition of relaxation times\cite{sasaki}. 
Our calculated results, while corroborating the findings of 
Ref.\cite{sasaki}, are based on a simpler model which also explains the 
suppression of this effect for the interacting sample. Assume that the 
system consists of single-domain particles of just two volumes $V_{1}$ and 
$V_{2}$. Recall that the 
characteristic relaxation time for a noninteracting single domain particle 
$\tau (V)\propto \exp[(KV \pm \mu HV)/k_{B}T]$, where K is the 
anisotropy energy constant, $\mu$ is the magnetic moment per unit volume, 
H the applied magnetic field and $k_{B}$ the Boltzmann constant. We have 
only two relaxation times $\tau_{1}$ and $\tau_{2}$ associated with $V_{1}$ 
and $V_{2}(V_{2}>V_{1})$. The time $\tau_{1}$ is much smaller than 
the measurement time while $\tau_{2}$ is much larger, at the lowest 
temperature $(T_{0})$ of measurement. 

\begin{figure}[t]
\begin{center}
\includegraphics[width=5.cm,angle=270]{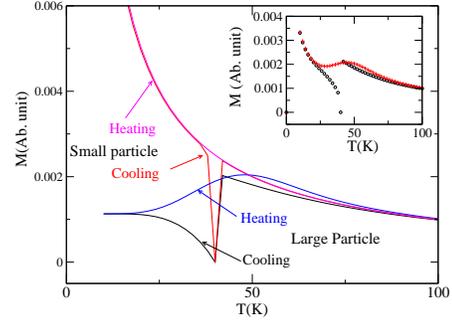}
\end{center}
\caption{Simulated arrested cooling and continuous heating for two different 
particle sizes. Inset shows the combined effect. }
\label{theory}
\end{figure}
Both $\tau_{1}$ and $\tau_{2}$ are 
expected to be smaller than the measurement time at the highest temperature 
$T_{\infty}$. Therefore, in the intermediate temperature domain 
$(T_{0} \le T \le T_{\infty})$, some of the particles will equilibriate 
rapidly, thus showing superparamagnetism while the others will be 'blocked'. 
This is observed in Fig.\ref{theory} where we have plotted M(T)
separately for the two sets of particles under the same cooling and heating 
protocol. When H is zero, both sets of particles relax to $M = 0$, however, 
when H is turned on, particles $2$ are blocked ($M = 0$) while $1$ show 
facile response. As T is increased again, M for $1$ decreases with T while 
M for $2$ {\em increases} thus producing a wiggle. 
This effect is seen only when the temperature of arrest is in-between the 
two respective blocking  temperatures. In the interacting system, one has 
an additional term $\propto V^2$ in the exponent of $\tau_V$ in a meanfield
picture which shifts both the blocking temperatures to higher T causing
the wiggles to disappear.  
\vskip .1cm

Discussions with P. A. Sreeram are gratefully acknowledged.

\noindent
S. Chakravarty$^1$, A. Frydman$^2$, M. Bandyopadhyay$^1$, S. Dattagupta$^1$ 
and S. Sengupta$^1$

\noindent
$^1$ S. N. Bose National Centre for Basic Sciences, Block JD, Sector III, 
Salt Lake, Kolkata 700 098 INDIA.\\
\noindent
$^2$ Dept. Of Physics, Bar Ilan University, Ramat Gan 52900, Israel.

\end{document}